\documentclass[prb,preprint,aps,superscriptaddress,longbibliography]{revtex4-1}

\usepackage{graphicx, epstopdf}
\usepackage{subfigure}
\usepackage{float}
\usepackage{amsmath}
\usepackage{amsfonts}
\usepackage{amssymb}
\usepackage{bm}
\usepackage{subfigure}
\usepackage{caption}
\usepackage{hyperref}
\begin{document}


\title{Quantum Spin Hall State on Square-like Lattice}


\author{Zhida Song}
\author{S. M. Nie}
\affiliation{Beijing National Laboratory for Condensed Matter Physics,
  and Institute of Physics, Chinese Academy of Sciences, Beijing
  100190, China}

\author{Hongming Weng}
 \email[]{hmweng@iphy.ac.cn}
 \affiliation{Beijing National Laboratory for Condensed Matter Physics,
  and Institute of Physics, Chinese Academy of Sciences, Beijing
  100190, China}
\affiliation{Collaborative Innovation Center of Quantum Matter,
  Beijing, 100084, China}
\author{Zhong Fang}
 \email[]{zfang@iphy.ac.cn}
 \affiliation{Beijing National Laboratory for Condensed Matter Physics,
  and Institute of Physics, Chinese Academy of Sciences, Beijing
  100190, China}
\affiliation{Collaborative Innovation Center of Quantum Matter,
  Beijing, 100084, China}


\date{\today}

\begin{abstract}
We find that quantum spin Hall (QSH) state can be obtained on a square-like or rectangular lattice,
which is generalized from two-dimensional (2D) transition metal dichalcogenide (TMD) haeckelites.
Band inversion is shown to be controled by hopping parameters and results in Dirac cones with opposite
or same vorticity when spin-orbit coupling (SOC) is not considered. Effective k$\cdot$p model has
been constructed to show the merging or annihilation of these Dirac cones, supplemented with the
intuitive pseudospin texture. Similar to graphene based honeycomb lattice system, the QSH insulator is
driven by SOC, which opens band gap at the Dirac cones. We employ the center evolution of hybrid Wannier
function from Wilson-loop method, as well as the direct integral of Berry curvature, to identify the $Z_2$ number. We hope our
detailed analysis will stimulate further efforts in searching for QSH insulators in square or rectangular lattice,
in addition to the graphene based honeycomb lattice.
\end{abstract}

\pacs{}

\maketitle

\section{Introduction}
Topological insulator (TI) has attracted significant attentions from researchers
in recent years.~\cite{hasan_colloquium:_2010, qi_topological_2011, bernevig_topo_2013, MRS_weng:9383312}
Its bulk band structure is characterized by a global topological invariant $\mathbb{Z}_2$ number,~\cite{kane_Z2_2005, bernevig_quantum_2006, fu_topological_2007, qi_general_2006, qi_topological_2008} which is protected by time-reversal (TR) symmetry. The bulk-boundary correspondence ensures that
on the boundary of a TI there exists topologically protected metallic states which cross the bulk band gap by connecting the valence and conduction bands.
Especially for two dimensional (2D) TI, the conducting one dimensional (1D) edge states have opposite velocity for opposite spin channel, which
leads to quantum spin Hall (QSH) effect. Therefore, 2D TI is also known as QSH insulator. Since the backscattering of electrons
in this 1D edge state is prohibited as long as TR symmetry is conserved, dissipationless quantized electron conductivity is expected and
has great potential application as an ideal conducting wire in device.

However, up to now, only two materials have been confirmed by experimental observation of quantized
conductivity inside of bulk band gap, namely, the quantum well structure composed of HgTe/CdTe~\cite{konig_quantum_2007}
and InAs/GaSb.~\cite{knez_evidence_2011} There have been intensive efforts in searching for 2D QSH insulator. One is based on
the honeycomb lattice, which is stimulated by Kane-Mele model for graphene.~\cite{kane_Z2_2005, kane_QSH_2005} The
low-buckled silicene~\cite{liu_quantum_2011}, chemically decorated single layer honeycomb lattice of
Sn,~\cite{PhysRevLett.111.136804} Ge~\cite{PhysRevB.89.115429} and Bi or Sb~\cite{BiX_Yao2014} are
all in this category. The second category is represented by
ZrTe$_5$/HfTe$_5$~\cite{weng_transition-metal_2014} and Bi$_4$Br$_4$~\cite{zhou_large-gap_2014}.
They are single layer exfoliated from their three-dimensional (3D) counterparts, which are weakly bonded layered materials.
The proposal of 1T' structure of single layer transition-metal dichalcogenide (TMD)~\cite{qian2014quantum} can be ascribed
to this type also. The third category are based on square or rectangular lattice. Such as recent proposals of 2D buckled square
lattice BiF~\cite{XiangSQ} and rectangular TMD haeckelites.~\cite{smnie2015, yanbh, daiying}


%

In this work, we will discuss the possible topologically nontrivial quantum states based on the model
built for 2D monolayer transition metal dichalcogenide (TMD) haeckelites,~\cite{PhysRevB.89.205402, terrones2014} among which MX$_2$ (M=Mo, W and X=S, Se, Te )
have been predicted to be QSH insulators by Nie {\it et al.}~\cite{smnie2015}, Sun {\it et al.}~\cite{yanbh} and Ma {\it et al}.~\cite{daiying} using first-principles calculations.


The paper is organized as follows.
In section \ref{derive},the effective tight-binding (TB) model of
    TMD haeckelites are constructed from symmetrical analysis.
In section \ref{symmetry}, the symmetry and topology of the lattice model and the topological
    phase transition have been discussed.
In section \ref{dirac_quadra}, the vorticity, merging or annihilation of Dirac cones have been demonstrated by k$\cdot$p model.
In section \ref{conclusions}, we provide a brief summary.

\section{Derivation of Tight Binding Model}\label{derive}
Monolayer of TMD in its top view, such as MoS$_2$, can be thought as a honeycomb lattice with Mo and S on the A and B sublattice site, respectively.
However, recently a kind of ordered square-octagonal distortion have been introduced in honeycomb TMD to form haeckelite structure~\cite{PhysRevB.89.205402, terrones2014}
as that in graphene. Such TMD haeckelites have pairs of square-octagon composed of M and X atoms, noted
as MX$_2$-4-8 to be distinguished from usual MX$_2$ with honeycomb lattice. In our previous work~\cite{smnie2015},
the first-principles calculation for WS$_2$-4-8 indicates the four bands around the Fermi level are mainly contributed
by $d_{z^2}$ orbitals of M atoms. The $p$ orbitals from X atoms are quite far from Fermi level and can be safely ignored. In this point of view,
the square-octagonal lattice can be further simplified to be a rectangular lattice with only four M atoms per unit cell as shown in Fig. ~\ref{bonds}.

\subsection{Hamiltonian without SOC Term}
We take $d_{z^2}$ orbital on each M atom as local orbital basis and note it as Wannier orbital
$w_n(\boldsymbol{r}-\boldsymbol{R}_l)$. Here $n$=1...4 indicates four nonequivalent M atoms
in one unit cell as shown in Fig.~\ref{bonds} and $\boldsymbol{R}_l$ is the lattice index.
Such rectangular lattice has space group $Pbam$ and the following symmetries are satisfied.


    \begin{description}
    \item[$\mathcal{P}$]
        inversion symmetry operator. M$_1$ (M$_2$) and M$_3$ (M$_4$) are related by $\mathcal{P}$.
    \item[$m_z$]
        mirror operator with mirror plane perpendicular to normal of the 2D lattice plane. All M atoms keep identity.
    \item[$ \left\{ C_{2x}| \frac{1}{2} \frac{1}{2} 0 \right\} $]
         This is a non-symmorphic operation. After $C_2$ rotation around $x$-axis (along a lattice), a glide
         of $(\frac{1}{2}, \frac{1}{2}, 0)$ (in unit of lattice vectors) is applied. This operation
         relates  M$_1$ (M$_3$) with M$_2$ (M$_4$).
    \end{description}

Therefore, there are only three independent hopping terms in tight-binding (TB) approximation and they are:
    \begin{eqnarray}
        t_1 = \langle w_1 | H_0 | w_2 \rangle \\
        t_2 = \langle w_1 | H_0 | w_{3t} \rangle \\
        t_3 = \langle w_1 | H_0 | w_4 \rangle
    \end{eqnarray}
Here $H_0$ is the Hamiltonian of system without considering of spin-orbit coupling (SOC). The
hopping parameters $t_1$, $t_2$ and $t_3$ are marked in Fig. \ref{bonds} also. All these parameters
must be real because $w_n$ and $H_0$ are real. After fourier transformation
    $H_{0nm}(\boldsymbol{k})=\sum_{\boldsymbol{R}} \langle w_n(0) | H_0
        | w_m(\boldsymbol{R}) \rangle e^{i\boldsymbol{k}\cdot\boldsymbol{R}}$,
    we get $\boldsymbol{k}$ dependent Hamiltonian $H_0(\boldsymbol{k})$ satisfying periodic
    condition $H_0(\boldsymbol{k})$=$H_0(\boldsymbol{k+G})$ with $\bf{G}$ being integer times of reciprocal lattice.
A $\bf{k}$-dependent gauge transformation $H_0(\boldsymbol{k}) \to U(\boldsymbol{k}) H_0(\boldsymbol{k}) U^\dagger(\boldsymbol{k})$
is applied to get a concise but not periodic form of $H_0(\boldsymbol{k})$ in Eq. (\ref{H0k}).
This form is more convenient for deriving the effective k$\cdot$p models used in section \ref{dirac_quadra}.
    \begin{align}
        H_0(\boldsymbol{k}) &=
        \left( \begin{array}{cccc}
             0 & \gamma_1 & \gamma_2^- & \gamma_3 \\
             \gamma_1^* & 0 & \gamma_3^* & \gamma_2^+ \\
             \gamma_2^{-*} & \gamma_3 & 0 & \gamma_1 \\
             \gamma_3^* & \gamma_2^{+*} & \gamma_1^* & 0
        \end{array}\right)  \label{H0k}\\
        \gamma_1 &= 2t_1\cos\left(\frac{k_x}{2}\right) \nonumber \\
        \gamma_2^\pm &=  t_2 e^{-i\frac{1}{2}(k_x \pm k_y)} \nonumber \\
        \gamma_3 &= 2t_3\cos\left(\frac{k_y}{2}\right) \nonumber
    \end{align}
    \begin{align}
        U(\boldsymbol{k}) = \left(
        \begin{array}{cccc}
            1 & 0 & 0 & 0 \\
            0 & e^{-\frac{i}{2}k_x} & 0 & 0 \\
            0 & 0 & e^{-\frac{i}{2}(k_x+k_y)} & 0 \\
            0 & 0 & 0 & e^{-\frac{i}{2}k_y}
        \end{array}
        \right)  \label{Uk}
    \end{align}

\subsection{Spin-Orbit Coupling Terms}
To include SOC term, the degree of freedom of spin is introduced explicitly to expand the dimension
of the Hamiltonian from four to eight. The spinor basis set is defined as:
    \begin{eqnarray}
        |\phi_i\rangle &= w_i(\boldsymbol{r}) \otimes \chi_\uparrow
            \quad i=1\cdots4 \\
        |\phi_{i+4}\rangle &= w_i(\boldsymbol{r}) \otimes \chi_\downarrow
            \quad i=1\cdots4
    \end{eqnarray}
, where $\chi_\uparrow$=$(1, 0)^T$ and $\chi_\downarrow$=$(0, 1)^T$ with $T$ representing transpose.
Correspondingly, all the above symmetrical operators should be adjusted:
    \begin{equation*}
        \mathcal{P} \to \mathcal{P} \\
    \end{equation*}
    \begin{equation*}
        m_z \to m_z e^{-\frac{i}{2}\pi\sigma_z} \\
    \end{equation*}
    \begin{equation*}
        \left\{C_{2x}|\frac{1}{2}\frac{1}{2}0\right\} \to
            \left\{C_{2x}|\frac{1}{2}\frac{1}{2}0\right\} e^{-\frac{i}{2}\pi\sigma_x}
    \end{equation*}
and TR operator is defined as $\mathcal{T} = -i\sigma_y K$ with K is complex conjugation operator.
Considering these symmetrical constraints, we will derive the term including SOC $H_{so}$.

{\it Mirror symmetry $m_z$.} Similar as the case in Kane-Mele model\cite{kane_QSH_2005} on honeycomb lattice without Rashba term,
$H_{so}$ for present TMD haeckelite is also block diagonal in spin space because of $m_z$ symmetry.
This can be proved as:
    \begin{align*}
        \forall i,j=1\cdots4 &\\
        \langle \phi_{i+4} | H_{so} | \phi_j \rangle &=
            \langle m_z e^{-\frac{i}{2}\pi\sigma_z} \phi_{i+4} | H_{so} |
            m_z e^{-\frac{i}{2}\pi\sigma_z} \phi_j \rangle \\
        &= - \langle \phi_{i+4} | H_{so} | \phi_j \rangle=0
    \end{align*}
The derivation from the second line to the third line is due to $e^{-\frac{i}{2}\pi\sigma_z}=-i\sigma_z$.

{\it Time-reversal symmetry $\mathcal{T}$.} $H_{so}^{\uparrow\uparrow}$ and
    $H_{so}^{\downarrow\downarrow}$ are related by $\mathcal{T}$:
    \begin{align}\label{Hso_cond1}
        \langle \phi_{i+4} | H_{so} | \phi_{j+4} \rangle
            & = \langle \mathcal{T} \phi_i | H_{so} | \mathcal{T} \phi_j \rangle \nonumber \\
            & = \langle \phi_j | H_{so} | \phi_i \rangle
    \end{align}
    i.e., $H_{so}^{\uparrow\uparrow} = H_{so}^{\downarrow\downarrow T}$.

\begin{figure}[]
\includegraphics[ angle=00, width=0.8\textwidth]{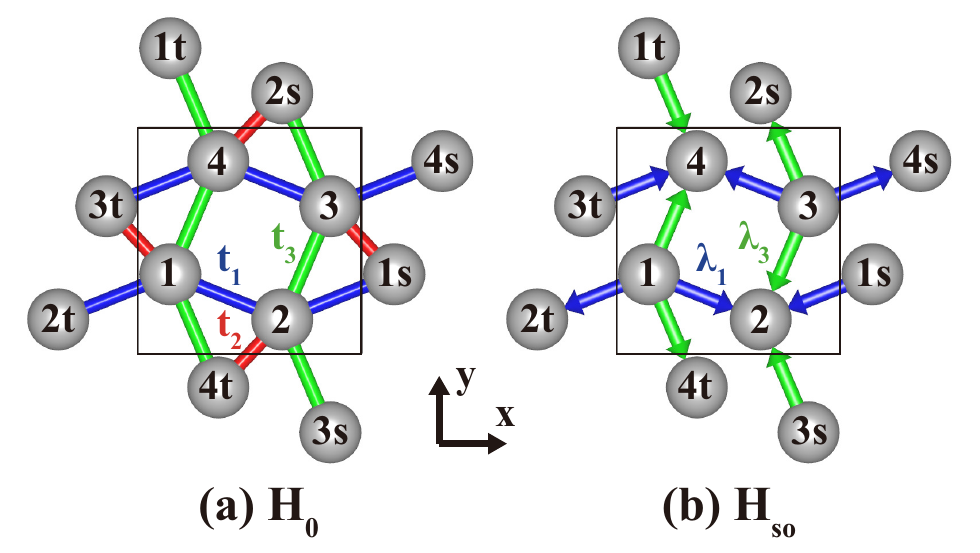}
\caption{2D rectangular lattice with four sties per cell. (a) Hopping parameters $t_1$, $t_2$ and $t_3$ in $H_0$.
  (b) Finite parameters $\lambda_1$ and $\lambda_3$ in $H_{so}$ with each arrow points from bra to ket.
    \label{bonds}}
\end{figure}

{\it Inversion symmetry $\mathcal{P}$ and combined symmetry $\left\{C_{2x}|\frac{1}{2}\frac{1}{2}0\right\}
    e^{-\frac{i}{2}\pi\sigma_x} \cdot \mathcal{T}$.}
These two operations do not flip spin. After an intuitive graphic analysis as illustrated
in Fig. \ref{bonds}, one can easily find that there are only two independent non-zero parameters:
    \begin{align}
     \langle \phi_1 | H_{so} | \phi_2 \rangle &=\langle \phi_3 | H_{so} | \phi_4 \rangle =i \lambda_1 \\
     \langle \phi_1 | H_{so} | \phi_4 \rangle &=\langle \phi_3 | H_{so} | \phi_2 \rangle =i \lambda_3
    \end{align}
 Note that these two SOC terms must be pure imaginary numbers because $\phi_i$
    are real and $H_{so}$ contains an imaginary unit. We also note that $\langle \phi_{1}| H_{so} | \phi_{3t} \rangle$
    must be zero because $\mathcal{P}$ transforms it to its complex conjugation.

Therefore, $H_{so}$ is always diagonal in spin space and $\mathcal{PT}$ guarantees spin degeneracy, i.e., Kramer degeneracy.
We only need to deal with $H_{so}^{\uparrow\uparrow}$ for the rest of this article.
The concise matrix form of $H_{so}$ in Eq. (\ref{Hsok}) is achieved by the
 same gauge transform (Eq. (\ref{Uk})) performed to one spin channel of $H_{so}(\boldsymbol{k})$.
    \begin{align}
        H_{so}^{\uparrow\uparrow}(\boldsymbol{k}) &= \left(
            \begin{array}{cccc}
            0  & \delta_1 & 0 & \delta_3 \\
            \delta_1^* & 0 & \delta_3^* & 0 \\
            0 & \delta_3 & 0 & \delta_1 \\
            \delta_3^* & 0 & \delta_1^* & 0
            \end{array}
        \right)  \label{Hsok}\\
        \delta_1 &= 2i\lambda_1 \cos \left( \frac{k_x}{2} \right) \nonumber \\
        \delta_3 &= 2i\lambda_3 \cos \left( \frac{k_y}{2} \right) \nonumber
    \end{align}

Now we get the total effective TB model  $H(\boldsymbol{k})$=$H_0(\boldsymbol{k})+H_{so}(\boldsymbol{k})$,
but keep in mind that the periodic condition is sacrificed to get the conciseness in
    Eq. (\ref{H0k}) and Eq. (\ref{Hsok}), and a gauge transformation (Eq. (\ref{Uk})) to periodic form is
    needed when calculating properties concerning global phase such as Berry phase and related properties.

\section{Symmetry and Topology of the Bands} \label{symmetry}
From the analysis of symmetry in the above section, we obtain a TB
 model as shown in Eq. (\ref{H0k}) and Eq. (\ref{Hsok}).
We now focus on the band structure and the topology with different choice of related parameters.
In real materials, these parameters are determined by details in crystal structure and choice of M and X elements
in MX$_2$-4-8 haeckelites.~\cite{smnie2015}
It should be noticed that only the sign($t_1 t_2 t_3$) makes sense.
This can be easily understood if we can use some unitary transformation $UHU^T$
to change $t_i$'s signs in pairs. For example, $U$=$\mathbb{I}_2 \otimes \tau_z$
     transforms $t_1$, $t_3$ to $-t_1$, $-t_3$ and $U$=$\tau_z \otimes \mathbb{I}_2$
     transforms $t_2$, $t_3$ to $-t_2$, $-t_3$.
The change in the $\text{sign}(t_1 t_2 t_3)$, whatever which sign($t_i$)
    changes, is equivalent to $H_0(\boldsymbol{k}) \to -H_0(\boldsymbol{k})$.
Without loss of generality, in the rest of this paper we deal with $0<t_3\le t_1$.
For the case $|t_1|\le |t_3|$, the system can be thought as the case $|t_1|\ge|t_3|$
    with the interchange of $x$ and $y$ axes, which is apparent in Fig. \ref{bonds}.
For clarity, we list all possible band structure with $0<t_3\le t_1$ in Fig. \ref{bandsall}.
The classification and topology of them will be discussed in the following.

\subsection{Symmetry of The Eigenstates}
\begin{table}
\caption{Eigen energies, their degeneracy and eigenvalues of symmetrical operations of
         $H_0(\boldsymbol{k})$ at four time-reversal invariant momenta (TRIM).
        \label{solution0}}
\begin{tabular}{|c|c|c|c|c|c|}
  \hline
  \hline
        & Energy & Folds & $\mathcal{P}$ & $\{C_{2x}|\boldsymbol{\tau}_x\}$ & $\{C_{2y}|\boldsymbol{\tau}_y\}$ \\
  \hline
    $\Gamma$ & $E_A^0 = t_2 -2(t_1+t_3) $ & 1 & 1 &-1 & -1 \\
    \cline{2-6}
             & $E_B^0 =-t_2-2(t_1-t_3) $ & 1 &-1 &-1 &  1 \\
    \cline{2-6}
             & $E_C^0 =-t_2+2(t_1-t_3) $ & 1 &-1 & 1 & -1 \\
    \cline{2-6}
             & $E_D^0 = t_2+2(t_1+t_3) $ & 1 & 1 & 1 &  1 \\
  \hline
    $X$     & $-\sqrt{t_2^2+4t_3^2}$ & 2 & $\pm 1$ & $\pm 1$ & $\pm 1$ \\
    \cline{2-6}
            & $ \sqrt{t_2^2+4t_3^2}$ & 2 & $\pm 1$ & $\pm 1$ & $\pm 1$ \\
  \hline
    $Y$     & $-\sqrt{t_2^2+4t_1^2}$ & 2 & $\pm 1$ & $\pm 1$  & $\pm 1$ \\
    \cline{2-6}
            & $ \sqrt{t_2^2+4t_1^2}$ & 2 & $\pm 1$ & $\pm 1$  & $\pm 1$ \\
  \hline
    $S$     & $ -t_2 $  & 2 & $1,1$ & $\pm i $   & $\pm i $  \\
    \cline{2-6}
            & $  t_2 $  & 2 & $-1,-1$ & $\pm i $   & $\pm i $ \\
  \hline
  \hline
\end{tabular}
\end{table}

\begin{figure*}[]
\includegraphics[ angle=00, width=0.8\textwidth]{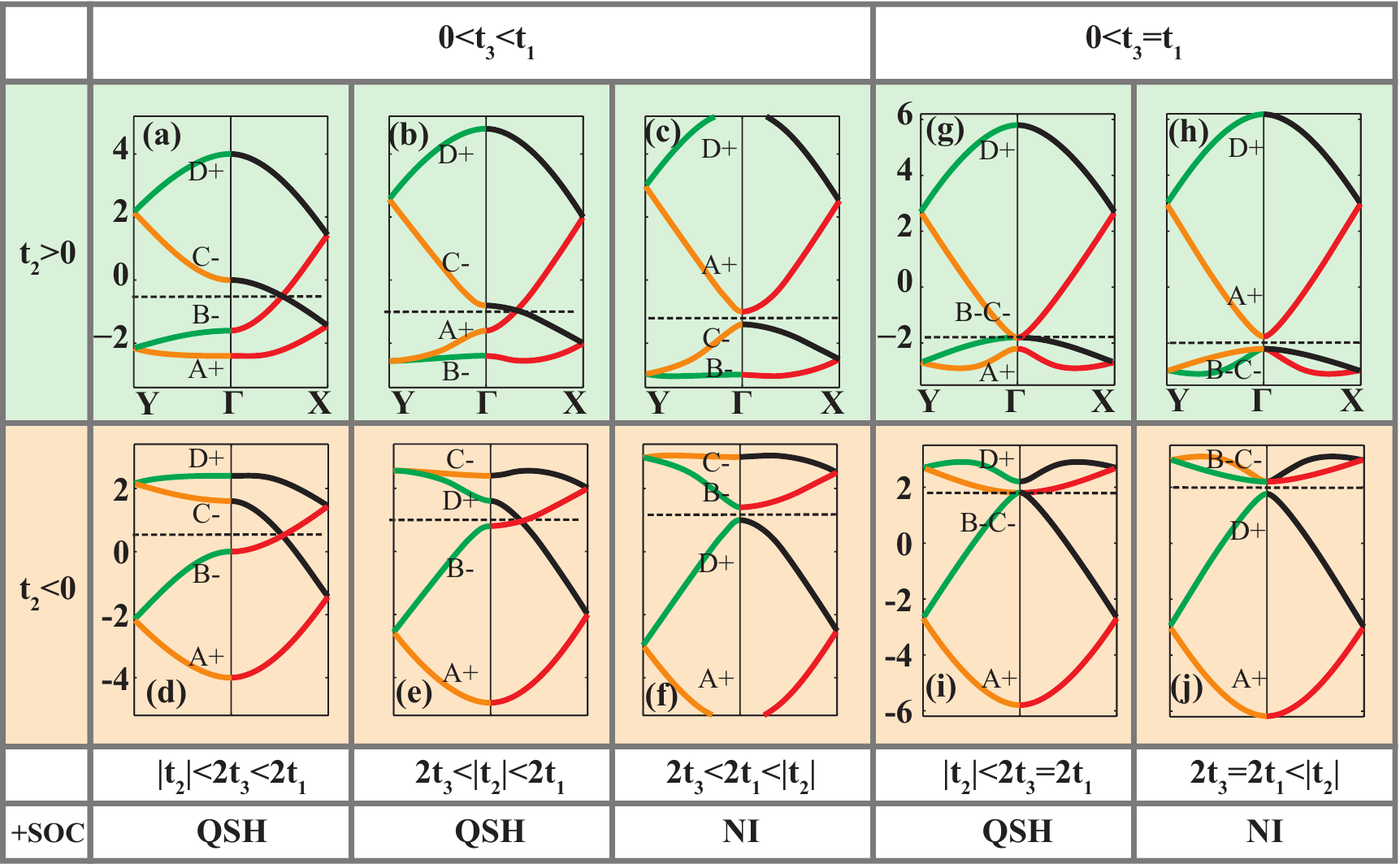}
\caption{Band structure of $H_0(\boldsymbol{k})$ for all possible parameter choices with $0<t_3\le t_1$.
(a)-(f) and (g)-(j) show bands without ($t_3\ne t_1$) and with ($t_3 = t_1$) $C_4$ symmetry, respectively. Upper and lower panels are
for $t_2>0$ and $t_2<0$ cases, respectively. Color of each band represents its representation under the
conserved non-symmorphic rotational symmetry. Along $\Gamma$-X ($\Gamma$-Y), black (green) and
red (yellow) represent +1 and -1 eigenvalue of $\{C_{2x}|\frac{1}{2}\frac{1}{2}0\}$ ($\{C_{2y}|\frac{1}{2}\frac{1}{2}0\}$), respectively.
Including SOC term will drive the system into QSH insulator or normal insulator (NI) as indicated for different parameter choice.
\label{bandsall}}
\end{figure*}

$H_0(\boldsymbol{k})$ at time-reversal invariant momenta (TRIM)~\cite{fu_topo_inv_2007} can be solved analytically.
The eigen energies, degeneracy and eigenvalue of symmetrical operators of the eigen states are listed in Table \ref{solution0}.
Little group at $\Gamma$ of $Pbam$ has only one-dimensional irreducible representations,\cite{bilbao2006}
    and all symmetrical operations can be diagonalized within the space spanned by eigen wavefunctions.

At TRIM $X(\pi, 0, 0)$ or $Y(0, \pi, 0)$, $Pbam$ has only 2D irreducible representations.
Within the eigen states of $H_0(\boldsymbol{k})$, they can be written as following:
    \begin{align}
        D_X\left(\mathcal{P}\right) = D_Y(\mathcal{P}) &= \begin{pmatrix}
                \tau_3 & 0 \\
                0 & \tau_3
            \end{pmatrix}  \\
        D_X\left(C_{2x}|\frac{1}{2}\frac{1}{2}0\right) &= 
            \begin{pmatrix}
                \tau_2 & 0 \\
                0 & \tau_2
            \end{pmatrix} \\
        D_Y\left(C_{2y}|\frac{1}{2}\frac{1}{2}0\right) &= 
            \begin{pmatrix}
                \tau_2 & 0 \\
                0 & \tau_2
            \end{pmatrix}
    \end{align}
    in which $\tau_i$ denotes the $i$-th pauli matrix, representing the space spanned by two-fold degenerate eigenstates
    of $H_0(\boldsymbol{X})$ or $H_0(\boldsymbol{Y})$.
    We can character symmetry properties of each eigen space by eigenvalues of symmetrical operators. For example, the two-fold degenerate eigenstates at either $X$ (or $Y$)
    have even and odd parities respectively since inversion symmetry $\mathcal{P}$ is represented as  $\tau_3$ (or $\tau_2$), thus we note this eigen space with $\pm1$ in Table \ref{solution0}.
    Solutions at $S (\pi,\pi,0)$ are special.
Little group of $Pbam$ at $S$ has only one-dimensional irreducible representations. However, for
    $\{C_{2x}|\frac{1}{2}\frac{1}{2}0\}$ and $\{C_{2y}|\frac{1}{2}\frac{1}{2}0\}$, the eigenvalues of the two eigenstates with the same parity
    are $\pm i$. Thus, the TR operator connects them and makes them two-fold degenerate.

From Table \ref{solution0}, we can see that the order of eigen energies at $X$ and $Y$ do not depend on the choice of $t_i$.
Further, since the parities of the double degenerate states are always in pair, the band inversion at $X$ and $Y$ will not change
the topology of the occupied bands. At $S$, the situation is similar, though the energy order depends on sign of $t_2$. Therefore,
only the band inversion at $\Gamma$ can result in different topology of occupied bands. This is similar to large-band gap 2D
topological insulator ZrTe$_5$.~\cite{weng_transition-metal_2014}

The order of energies at $\Gamma$ depends on $t_i$ explicitly as illustrated in Fig. \ref{bandsall}.
Along $\Gamma$-X(Y), $\{C_{2x}|\frac{1}{2}\frac{1}{2}0\}$ ($\{C_{2y}|\frac{1}{2}\frac{1}{2}0\}$)
symmetry is conserved and the eigenstates with the same eigenvalue of $\{C_{2x}|\frac{1}{2}\frac{1}{2}0\}$
($\{C_{2y}|\frac{1}{2}\frac{1}{2}0\}$) at different
k points form a continuous band, which is identified by different color in Fig. \ref{bandsall}.
As long as $|t_2|<2t_1$, the $\{C_{2x}|\frac{1}{2}\frac{1}{2}0\}$ eigenvalues of the two lower bands
at $\Gamma$ are $-1, -1$, while those at $X$ are $1, -1$. There must be a level crossing along $\Gamma$-X path.
However, if $|t_2|>2t_1$ , the eigenvalues of the two lower bands at
    $\Gamma$ and $X$ are the same and so there is no crossing. But we'd like to remind that even without the above $C_2$ symmetry,
    as long as TR and $\mathcal{P}$ is conserved, such spineless 2D lattice model will have stable band crossing once band inversion
    happens.~\cite{allcarbon, Cu3NPd} $C_2$ symmetry constraints the crossing point on the path along $\Gamma$-X or $\Gamma$-Y, which is
    similar to the spinless graphene model.\cite{bernevig_topo_2013} A detailed discussion about these crossing points will be carried
    out in section \ref{dirac_quadra}.

%

\subsection{Topological Phase}\label{topo_phase}
\begin{figure}[]
\includegraphics[ angle=00, width=0.8\textwidth]{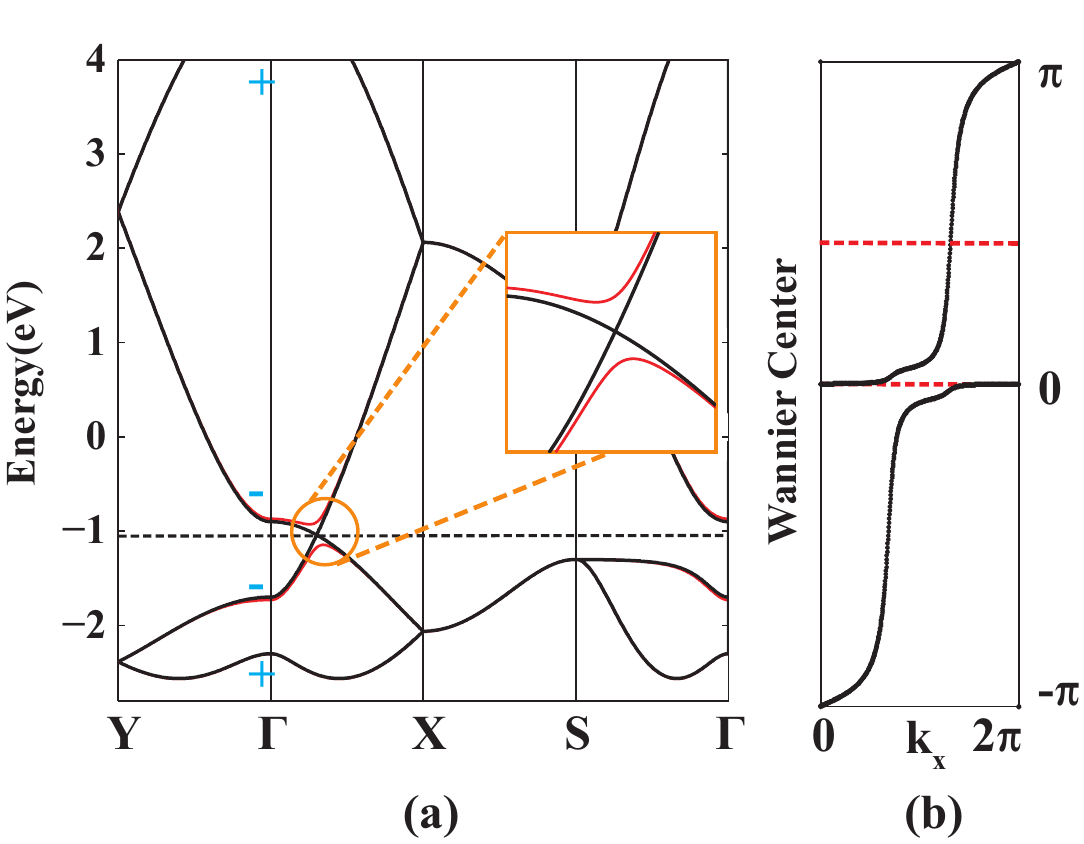}
\caption{Bands and wannier centers' evolution with SOC. (a) shows bands in which
         black lines denote non-SOC bands and red lines denote SOC bands.
         (b) shows wannier centers' evolution of spin up electrons.
         The dashed red lines indicate reference lines.
         The parameters are set as $t_1=1$, $t_2=1.3$, $t_3=0.8$, $\lambda_1 = -\lambda_3=0.04$.
    \label{bands_wilson}}
\end{figure}

We now investigate whether a QSH insulating state can occur or not after SOC term is considered,
which is similar to what happens in Kane-Mele model.\cite{kane_QSH_2005,kane_Z2_2005}
The $\mathbb{Z}_2$ number of system with both TR
and inversion $\mathcal{P}$ symmetries is determined by the product of parities of occupied states at four TRIM.\cite{fu_topo_inv_2007}
SOC term opens band gap at the band crossing points as shown in Fig. \ref{bands_wilson}(a).
As long as $|t_2|<2 \text{max}(t_1,t_3)$, $\mathbb{Z}_2$ number is 1 as shown in Fig.~\ref{bandsall}.
To verify this we calculate the evolution of hybrid Wannier function centers for occupied bands using
wilson loop method.\cite{yu_wilsonloop_2011, MRS_weng:9383312} The plot is shown in Fig. \ref{bands_wilson}(b).
This illustrates that the winding number of two occupied bands is 1, indicating that the Chern number of spin
up bands is 1. Due to TR symmetry, the spin down bands should have Chern number $-1$.
Therefore, the model with $|t_2|<2 \text{max}(t_1,t_3)$ and non-zero SOC parameter $\lambda_i$ is a QSH insulator state.
For the case $|t_2|>2 \text{max}(t_1,t_3)$, there will be no band inversion and the product of parities is 1, i.e., the
system becomes a trivial normal insulator (NI).

Explicitly, we investigate the topological phase transition (TPT) between TI and NI driven by tuning one parameter $t_2$.
Such TPT from QSH insulator to NI in centrosymmetric system will experience a gapless state with a Dirac
node at TRIM.~\cite{murakami_phase_2008, murakami_phase_2011} 
As illustrated in Fig.~\ref{bandsall}, the TPT can be thought as the bands inversion of $A$ and $C$
    at $\Gamma$ in case $t_2 >0$ or $B$ and $D$ in case $ t_2 <0$.
Hence, at TPT point the eigen energies of $H_0(\Gamma)+H_{so}(\Gamma)$ satisfy $E_A=E_C$ or $E_B=E_D$ with
    \begin{align}
    E_A &= t_2 - 2\sqrt{(t_1+t_3)^2+(\lambda_1+\lambda_3)^2} \\
    E_B &=-t_2 - 2\sqrt{(t_1-t_3)^2+(\lambda_1-\lambda_3)^2} \\
    E_C &=-t_2 + 2\sqrt{(t_1-t_3)^2+(\lambda_1-\lambda_3)^2} \\
    E_D &= t_2 + 2\sqrt{(t_1+t_3)^2+(\lambda_1+\lambda_3)^2}.
    \end{align}
All of the corresponding eigen states are spin degenerate.
It's easy to find the critical value of $t_2$:
    \begin{align}
    t_{2C} = & \pm\sqrt{(t_1+t_3)^2+(\lambda_1+\lambda_3)^2} \nonumber \\
             & \pm\sqrt{(t_1-t_3)^2+(\lambda_1-\lambda_3)^2} \label{t2c}
    \end{align}
Since $t_{2C}$ depends on $t_i$ and $\lambda_i$ ($i$=1 and 3),  band inversion is influenced by both hopping effect and SOC in real materials.

\section{Dirac points and quadratic band touching}\label{dirac_quadra}
From the above discussion, our model can be looked as an analog of graphene on a rectangular lattice: a significant resemblance is that QSH state
occurs when SOC term opens gap at Dirac points. Since the lattice constants $a$ and $b$ are quite close to each other in previously studied TMD
haeckelites,~\cite{smnie2015} we'd like to go further to see what will happen in a square lattice with $C_4$ symmetry, which is the case studied
in Ref.~\onlinecite{PhysRevB.89.205402} and ~\onlinecite{daiying}.
We find that in square lattice the non-SOC band structure has a quadratic band touching at $\Gamma$, being different from the linear
Dirac cones in graphene and rectangular lattice. As all interesting physics are related to the linear Dirac cone or the quadratic non-Dirac
cone band touching, in this section we use effective k$\cdot$p model to characterize them.
Alternatively, instead of calculating the evolution of centers of hybrid Wannier functions,~\cite{yu_wilsonloop_2011,MRS_weng:9383312}
we calculate the integral of berry curvature analytically by treating $H_{so}$ as an infinity small perturbation term to identify the
topological invariant.

\begin{figure*}[]
\centering
\includegraphics[ angle=00, width=0.8\textwidth]{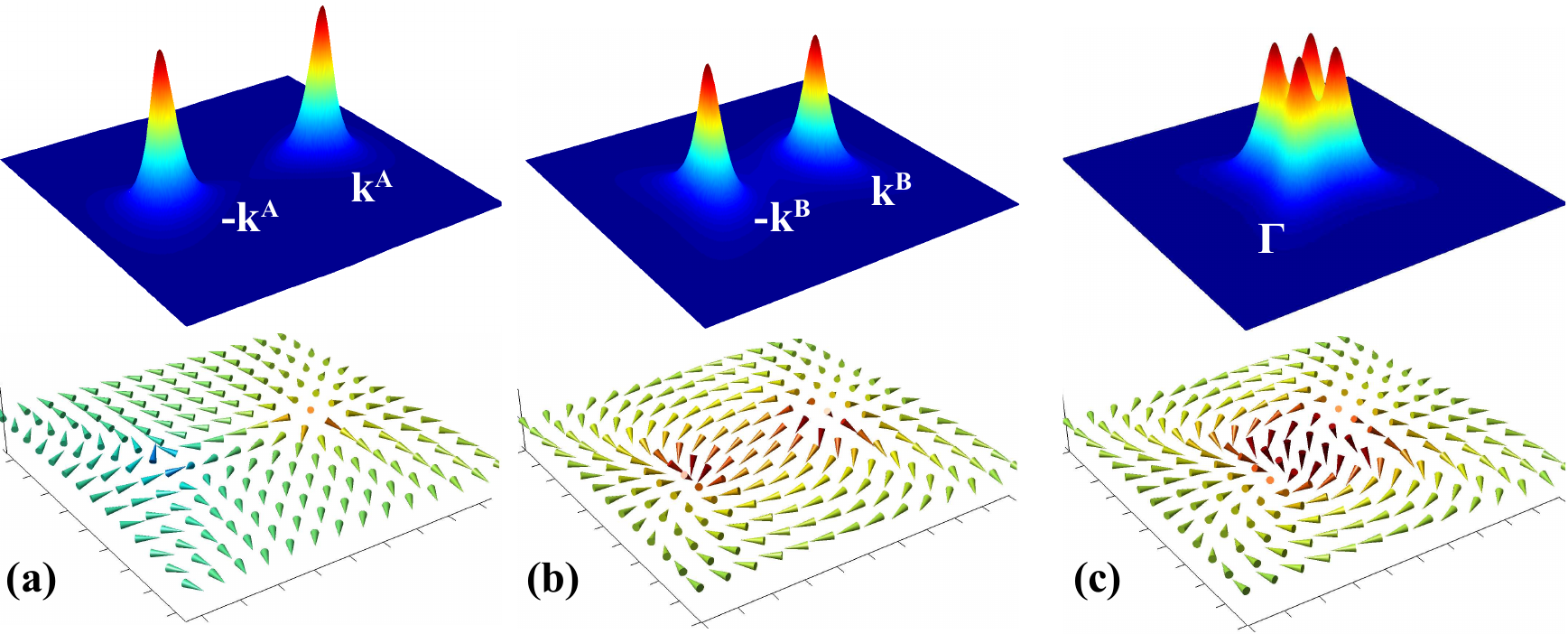}
\caption{Berry curvature and pseudospin distribution in momentum space around around band crossing or touching points.
        (a) shows the Dirac cones with opposite vorticity. (b) and (c) show the Dirac cones having the same vorticity
        when $C_4$ symmetry is slightly broken and conserved , respectively.
    \label{dirac}}
\end{figure*}

\subsection{Linear Dirac Nodes in Rectangular lattice}
As the non-trivial topology comes from band inversion of A, C or B, D bands around $\Gamma$,
it's natural to choose the states of A, C or B, D as the bases to establish k$\cdot$p
model to capture the essence of the underlying physics. We take the case of A, C band inversion in the following
discussion and the B and D inversion case can be easily inferred from it.
We expand Eq. (\ref{H0k}) and Eq. (\ref{Hsok}) to second order of $\boldsymbol{k}$ and then
    use the downfolding technique\cite{voon_k_2009,winkler_spin-orbit_2003} to reduce the dimensionality
   in order to get the effective Hamiltonian in the bases of eigenstates $|A\rangle$ and $|C\rangle$.
To keep the k$\cdot$p model similar to the Kane-Mele model, we transform $|A\rangle$ and $|C\rangle$ to the following
two bases:
    \begin{align}
    |1\rangle = \frac{-|w_1\rangle+|w_4\rangle}{\sqrt{2}} \qquad
    |2\rangle = \frac{ |w_2\rangle-|w_3\rangle}{\sqrt{2}}
   \end{align}
 Within the above basis set, the space inversion and TR operators are
 \begin{align}
  D_{AC}(\mathcal{P}) = \tau_1 \quad & \text{and}~~~D_{AC}(K) =  K  \label{DAC}
 \end{align}
 , respectively, both of which are the same as those in Kane-Mele model.

After some approximation (neglecting second order of $\lambda_i$), we get the k$\cdot$p model in
basis set of $|1\rangle$ and $|2\rangle$:
    \begin{widetext}
    \begin{align}
        H_{AC}(\boldsymbol{k}) &= E_{AC}(\boldsymbol{k})
            + \left[ t_2 \left( 2-\frac{1}{8}\boldsymbol{k}^2 \right) - t_1 \left( 2-\frac{1}{4}k_x^2 \right)
                      + F k_x^2 \right] \tau_1
            -\frac{t_2}{2}k_y \tau_2 + J k_x \tau_3 \label{HAC}\\
        F &= \frac{t_2^2}{8(E_A-E_B)} - \frac{t_2^2}{8(E_C-E_D)} \\
        J &=-\frac{t_2}{4}\left[
            \frac{\Delta_-}{E_A-E_B} + \frac{\Delta_-}{E_C-E_B} + \frac{\Delta_+}{E_A-E_D}
            +\frac{\Delta_+}{E_C-E_D}
            \right]
    \end{align}
    \end{widetext}
where $E_{AC}(\boldsymbol{k})$ is an even function of $\boldsymbol{k}$ and $\Delta_\pm$
    is defined as $\Delta_\pm=2(\lambda_1\pm \lambda_3)$.

If SOC is not present, i.e., $\Delta_\pm$=0, there will be two Dirac cones from Eq.~\ref{HAC} and the location
of the two nodes $\pm\boldsymbol{k}^A$ are determined by solution of $H_{AC}(\boldsymbol{k})=0$. $\boldsymbol{k}^A$ is found to be:
    \begin{align}
    \boldsymbol{k}^A &= \boldsymbol{e}_1\sqrt{\frac{2t_1-t_2}{F+(2t_1-t_2)/8}} \label{kC}.
    \end{align}
$\boldsymbol{e}_1$ is the unit vector along $\Gamma$-X path. In the case $t_1<t_3$, $\boldsymbol{k}^A$ is along $\Gamma$-Y.
Expanding the hamiltonian Eq.~\ref{HAC} around $\boldsymbol{k}^A$ or $-\boldsymbol{k}^A$ to the lowest order of $\Delta_\pm$
and $\boldsymbol{q}$ (here a general $\boldsymbol{k}$ is $\pm\boldsymbol{k}^A$+$\boldsymbol{q}$), we get a linear dispersion hamiltonian, i.e. a Dirac hamiltonian:
    \begin{align}
        H(\pm\boldsymbol{k}^A + \boldsymbol{q}) &=
            \pm\alpha k_1^A q_1 \tau_1 - \frac{t_2}{2}q_2\tau_2
            \pm J k_1^A \tau_3  \label{H_AC} =\boldsymbol{d}\cdot\boldsymbol{\tau} \\
        \alpha &= \frac{2t_1-t_2}{4}+2F
    \end{align}
Here $\boldsymbol{d}$=$(d_1,d_2,d_3)$ and $\boldsymbol{\tau}$=$(\tau_1,\tau_2,\tau_3)$.
Representation of $\mathcal{P}K$ is $\tau_1 K$ and $(\mathcal{P}K)^{-1} H(\boldsymbol{k}) \mathcal{P}K =  H(\boldsymbol{k})$
guarantees the mass term (the term proportional to $\tau_3$ ) to be zero. Therefore, $\boldsymbol{d}$ becomes a 2-component vector. The two Dirac cones
possess opposite vorticity defined as $\mathrm{sign}(\det( {\partial \boldsymbol{d}}/{\partial \boldsymbol{q}}))$. In this sense, any perturbation holds $\mathcal{P}K$ symmetry can only move
the positions of  Dirac cones but can not annihilate them. This can also be easily seen from the texture of pseudospin, which is defined by
pseudospin operator $\hat{\boldsymbol{\tau}}$, As shown in Fig. \ref{dirac}(a), the pseudospin of occupied state rotating $2\pi$ along
an anticlockwise loop surrounding a Dirac cone gives a positive vorticity, while $-2\pi$ gives a negative vorticity. Pseudospin rotates 0
along a loop containing the two Dirac cones and a trivial insulating state will be obtained when these two Dirac cones meet together.

In the presence of SOC term, the time reversal operator becomes $-i\sigma_y K$ which is different from $K$ in spinless case.
Hence complex conjugation symmetry is broken and a mass term is introduced which drives the system to QSH state as discussed above.
Using the formula\cite{qi_general_2006,qi_topological_2011}:
    \begin{equation}
       \mathcal{F} = \frac{1}{2} \hat{\boldsymbol{d}} \cdot
            ( \partial_{k_x} \hat{\boldsymbol{d}} \times \partial_{k_y} \hat{\boldsymbol{d}} )
            \label{Fd}
    \end{equation}
where $\hat{\boldsymbol{d}}=\boldsymbol{d}/|\boldsymbol{d}|$, we calculate the berry curvature $\mathcal{F}$ around Dirac nodes:
    \begin{align}
        \mathcal{F}_{AC}(\boldsymbol{q}) &=
        \frac{\alpha t_2 J {k_1^{A}}^{2}}
            {4\left( \alpha^2 {k_1^{A}}^{2} q_1^2 + t_2^2 q_2^2 /4 + J^2 {k_1^{A}}^{2} \right) ^\frac{3}{2}}
            \label{curv_AC}
    \end{align}
Integral of Eq. (\ref{curv_AC}) around one Dirac point gives $\frac{1}{2}\mathrm{sign}({\alpha}t_2J)$, thus
the Chern number in spin up channel is $1$ (or $-1$). Due to TR symmetry, spin Chern number would be $1$ (or $-1$).


\subsection{Quadratic Band Touching in Square Lattice with $C_4$ Symmetry}
In this subsection, we will consider the square lattice with $C_4$ symmetry. This leads to $t_3=t_1$ and $\lambda_3=-\lambda_1$.
Firstly, we consider $H_0(\boldsymbol{k})$ without SOC. As shown in Fig.~\ref{bandsall}g and ~\ref{bandsall}i,
there is gapless band touching at $\Gamma$ and it is shown to be quadratic in the following discussion. This is due
to the degeneracy of $E_B$ and $E_C$ at $\Gamma$ and band inversion of $E_A$ and $E_C$.
If band inversion is absent, it becomes a trivial normal insulator as shown in Fig.~\ref{bandsall}h and~\ref{bandsall}j. These
two cases are similar to zinc-blend HgTe with band inversion and CdTe without band inversion.~\cite{bernevig_science_quantum_2006}

Although the non-trivial topology comes from the A, C band inversion, the k$\cdot$p model
based on bands A and C can not reproduce the band touching at Fermi level because it comes from the degeneracy of B and C bands.
Therefore, we construct a k$\cdot$p model in bases of $|B\rangle$ and $|C\rangle$. To keep the similar
form as $H_{AC}$, we also transform $|B\rangle$ and $|C\rangle$ to two new bases:
    \begin{align}
        |1\rangle &= \frac{|w_1\rangle-i|w_2\rangle-|w_3\rangle+i|w_4\rangle}{2} \\
        |2\rangle &= \frac{i|w_1\rangle-|w_2\rangle-i|w_3\rangle+|w_4\rangle}{2}
    \end{align}
Thus, the space inversion and complex conjugation operators are written as:
    \begin{align}
    D_{BC}(\mathcal{P}) = -1\qquad D_{BC}(K) = -i\tau_1 K
    \end{align}
which is different from Eq.~\ref{DAC} and is responsible for the quadratic band touching.
We get the Hamiltonian defined in basis set consisting of $|1\rangle$ and $|2\rangle$:
    \begin{widetext}
    \begin{align}
        H_{BC}(\boldsymbol{k}) &= -t_2 \left(1-\frac{1}{8}\boldsymbol{k}^2 \right)
            +\left[ t_3 \left( 2-\frac{1}{4}k_y^2 \right)
                    -t_1 \left( 2-\frac{1}{4}k_x^2 \right) \right]\tau_1
            +\frac{t_2}{4}k_x k_y \tau_2 + \Delta_- \tau_3  \label{HBC}
    \end{align}
    \end{widetext}

Similar to the last subsection, by expanding the Hamiltonian around band touching points $\pm\boldsymbol{k}^B$ to
the lowest order of $\Delta_\pm$ and $\boldsymbol{q}$, we get two Dirac cones:
    \begin{align}
            H_{BC}(\pm\boldsymbol{k}^B + \boldsymbol{q}) &=
            \pm\frac{t_1 k_{1}^B}{2} q_1  \tau_1
            \pm\frac{t_2 k_{1}^B}{4} q_2  \tau_2 + \Delta_-  \tau_3  \label{H_BC}\\
            \boldsymbol{k}^B &= \boldsymbol{e}_1 \sqrt{8-8\frac{t_3}{t_1}}  \label{kB}
    \end{align}
If $C_4$ symmetry is slightly broken, the two linear Dirac cones are slightly separated
by $2|\boldsymbol{k}^B|$. They are related by space inversion. However, because $\mathcal{P}=-1$ the two Dirac cones have
the same vorticity. The pseudospin texture in Fig.~\ref{dirac}(b) indicates the two Dirac cones have the same vorticity since
pseudospin rotates $4\pi$ along an anticlockwise loop surrounding the two Dirac cones. This is the reason of quadratic band touching
at $\Gamma$ when $C_4$ symmetry brings these two Dirac cones together. Such quadratic band touching can also be seen from
Eq.~\ref{HBC} by taking $t_3=t_1$ and $\lambda_3=-\lambda_1$ required by $C_4$ symmetry:
\begin{equation}
\small
H_{BC}(\boldsymbol{k}) = E_{BC}(\boldsymbol{k}) +\frac{t_1}{4}(k_x^2-k_y^2)\tau_1
    + \frac{t_2}{4}k_xk_y \tau_2 + \Delta_- \tau_3
\end{equation}
where $E_{BC}(\boldsymbol{k})$ is an even function of $\boldsymbol{k}$.

Similarly, SOC breaks complex conjugation symmetry and introduces a non-zero mass
and removes the above quadratic band touching degeneracy at $\Gamma$. This leads the system into insulator.
To identify the band topology of this insulating state, the momentum distribution of berry curvature
around $\Gamma$ can be calculated by Eq. (\ref{Fd}):
\begin{align}
    \mathcal{F} &= \frac{t_1t_2\Delta_- k^2}
    {4\sqrt{16\Delta_-^2 + k^4\left[t_1^2\cos^2(2\theta)+\frac{t_2^2}{4}\sin^2(2\theta)\right]}} \\
    \theta &= \text{arccos}\left(\frac{k_x}{|\boldsymbol{k}|}\right) \nonumber
\end{align}
It is shown in Fig. 4(c). The four-fold symmetrical distribution can be easily seen and understood
from the $2\theta$ terms. The integral of this berry curvature gives 1 (or $-1$). This indicates the Chern number
in spin up channel is 1 (or $-1$) and the time-reversal symmetry ensures this insulating state to be a QSH insulator.
As shown in Fig.~\ref{dirac}(c), the two Dirac cones are now moved together to $\Gamma$ to form a double
Dirac cone,~\cite{HgCrSe} and the pseudospin rotates $4\pi$ after adiabatic loop evolution surrounding
it,~\cite{HgCrSe} which indicates the quadratic band touching is non-trivial after $H_{SO}$ is included.

\section{Conclusions} \label{conclusions}
In this paper, we derived an low energy TB model for 2D TMD haeckelites MX$_2$-4-8 to uncover the physical mechanism controlling
the topological phase transition. This square-like or rectangle lattice model contains only one orbital per site and four sites per unit cell.
The band inversion is tuned by hopping parameters, which brings linear Dirac cones with the same or opposite vorticity. The annihilation
or merging of them has been demonstrated through effective k$\cdot$p model analysis and elaborated by pseudospin texture. SOC plays
a critical role in opening band gap at the band crossing or band touching points, which drives the system into insulator. The hybrid Wannier
function center evolution calculated by Wilson loop method, as well as the direct integral of Berry curvature, is used to identify the topology of
insulating state. The QSH insulator can be obtained within reasonable parameter region and can be used to understand the band
structures of TMD haeckelites MX$_2$-4-8 with M=Mo, W and X=S, Se, Te. Our model has extended the QSH lattice model
family to square or rectangular lattice.


\begin{acknowledgments}
We acknowledge the supports from National Natural Science Foundation of China (No. 11274359 and 11422428),
the 973 program of China (No. 2011CBA00108 and 2013CB921700) and the ``Strategic Priority Research Program (B)"
of the Chinese Academy of Sciences (No. XDB07020100). Part of the calculations were preformed on TianHe-1(A), the
National Supercomputer Center in Tianjin, China.
\end{acknowledgments}

\bibliography{WTe2-Model}

\end{document}